\input harvmac

\def\np#1#2#3{Nucl. Phys. {\bf B#1} (#2) #3}

\def\pl#1#2#3{Phys. Lett. {\bf #1B} (#2) #3}
\def\prl#1#2#3{Phys. Rev. Lett.{\bf #1} (#2) #3}

\def\cmp#1#2#3{Comm. Math. Phys. {\bf #1} (#2) #3}


\lref\dpol{ J. Polchinski, \prl {75} {1995} {4724} }  
\lref\BFSS{T. Banks, W. Fischler, S. H. Shenker and L. Susskind,
   {\it hepth} {9610043}}
\lref\witbd{ E. Witten, {\it hepth} {9510135}}
\lref\bsv{ M. Bershadsky, V. Sadov, C. Vafa, \np {463}{1996}{420} } 
\lref\senmarg{ A. Sen {\it hepth} {9510229} }
\lref\thtor{ 't Hooft, \cmp{81} { 1981} {267}  }
\lref\ambfly{ J. Ambjorn and H. Flyberg, \pl {97}  {1980} 241 }
\lref\susmal{ J. Maldacena and L. Susskind, \np {475} {1996} {679} } 
\lref\hamo{ J. Harvey and G. Moore, { \it hepth} 9609017  } 
\lref\BL{ V. Balasubramanian and R. Leigh, {\it hepth} 9611165.   } 
\lref\BDL{ M. Berkooz, M. Douglas, R.Leigh, 
             {\it  hepth} 9606139, \np {480}{1996}{265}  } 
\lref\gilad{ G. Lifschytz, {\it hepth }  9610125 } 
\lref\BMM{ J.Breckenridge, 
             G. Michaud, R. Myers, {\it hepth} 9611174. } 
\lref\malsus{J. Maldacena,  L. Susskind,   {\it hepth} \np {475} {1995} {679}}
\lref\mal{ J. Maldacena, \np {477} {1996} {168} }
\lref\cgkt{ C. Callan, S. Gubser, I. Klebanov, A. Tseytlin, 
           {\it hepth} 9610172.  } 
\lref\malstrom{ J. Maldacena, A. Strominger, {\it hepth} 9609026}
\lref\km{ I. Klebanov, S. Mathur, {\it hepth} 9701187}  
\lref\dm{ S. Das, S. Mathur, {\it hepth} 9607149, \np {482} {1996} {153} }
\lref\akfrac{ A. Hashimoto, {\it hepth } 9610250. }
\lref\malstromii{ J. Maldacena, A. Strominger, {\it hepth} 9702015 }
\lref\wati{ W. Taylor IV, {\it hepth} 9611042. } 
\lref\sus{ L. Susskind,   {\it hepth} 9611164. }    
\lref\grt{ O. J. Ganor, S. Ramgoolam, W. Taylor IV, {\it hepth} 9611202} 
\lref\domo{ M. Douglas and G. Moore, {\it hepth} 9603167. } 
\lref\bss{ T. Banks, N. Seiberg, S. Shenker, {\it hepth} 9612157.  } 
\lref\gil{ G. Lifshytz, {\it hepth} 9612223 } 
\lref\docoup{ M. Douglas, {\it hepth} 9604198 } 
\lref\polrev{ J. Polchinski, {\it hepth}  9611050. }



\Title{\vbox{\baselineskip12pt\hbox{hep-th/9702099, PUPT-1683 }}}
{\vbox{
\centerline{ Torons and D-brane bound states.}}}
\centerline{Z. Guralnik and S. Ramgoolam. }
\smallskip
\smallskip
\centerline{Department of Physics, Jadwin Hall}
\centerline{Princeton University}
\centerline{Princeton, NJ 08544, USA}
\centerline{\tt zack,ramgoola@puhep1.princeton.edu}
\bigskip
\bigskip
\bigskip
\noindent
We interpret instantons on a torus 
with  twisted boundary conditions, in terms of 
bound states of branes. The interplay between 
the $SU(N)$ and $U(1)$ parts of the $U(N)$ theory  
of $N$ 4-branes allows the construction of 
a variety of bound states.  The $SU(N)$ and $U(1)$ 
parts can contribute fractional amounts to 
the total instanton number which is integral.  
The geometry of non-self intersecting two-cycles 
in $T^4$ sheds some light on a number of 
properties of these solutions.

\Date{ February, 1997}

\newsec{ Introduction}

Recent developments in string duality
 have shown that many aspects 
of gauge theories can be understood geometrically 
by using the fact that the  low energy description of D-branes
is Yang Mills theory \dpol\witbd. 
Here we explore another aspect of this 
geometrization of Yang Mills by interpreting torons
 in terms of bound states of D-branes.

Torons are instantons on a torus \thtor, possibly 
 with fractional instanton number, 
 which exist when we impose twisted boundary conditions.
They were studied by 't Hooft in $SU(N)/Z_N$ gauge theory.  
 We discuss their embedding in  $U(N)$ gauge theories, and 
interpret  in terms 
 of bound states of 4-branes, 2-branes and 0-branes. 
The fact that the global structure of the gauge group 
needed for the D-branes is   $U(N)$, as opposed to 
$SU(N)\times U(1)$,  will be crucial.  
 In this paper our considerations will be mostly 
 classical. We consider  these solutions in the Euclidean $T^4$ of  
 supersymmetric 4+1 dimensional Yang Mills (with $16$ supersymmetries) 
 on a torus $T^4 \times R$, and interpret in terms 
of  4-branes which have the four 
spatial dimensions along the  $T^4$. 
At this point we should comment on the interest in studying 
the classical solutions. The correct counting of quantum states 
in the sector specified by a set of brane charges is 
given by solving the quantum mechanics on the moduli 
space. This can often be reformulated in terms of 
the cohomology of an appropriate compactification of the moduli 
space. Superficially different but related cohomological 
formulations have been given in \bsv\ and \hamo. 
 From this point 
of view, this paper is largely a   
  step in investigating parts of these 
 moduli spaces which can be realized simply in terms 
of fractional instantons. Placing fractional instantons
in the D-brane context in this way leads to some 
simple geometric insights
into the conditions of their existence. 

The simplest class of bound states we consider will 
be denoted by $(4220)$. These 
have $4$-brane charge corresponding to  
4-branes  wrapped around the compact spatial directions
$1,2,3,4$;  $2$-brane charge corresponding to the 
 a $2$-branes wrapped along the directions $(12)$; 
 $2$-brane charge corresponding to $2$-branes 
wrapped along the $(34)$ direction;  and zero brane charge.
Such bound states have been discussed from the string theory 
point of view in \gilad\ and from supergravity in \BMM.
In the context of the M(atrix)  
models of Banks, Fischler, Shenker and Susskind \BFSS, systems
 with these charges have also been discussed recently \grt\bss\gil.   

The next class of solutions we 
consider   has charges $(422)$. These 
manage to have a vanishing zero brane charge 
thanks to a cancelation of the  fractional instanton number 
between the $SU(N)$ and the $U(1)$ parts of the 
theory. They  also manage to be 
  supersymmetric configurations of the D-brane theory
although they are not susy-configurations if we take the susy 
transformations to be those of Yang Mills alone. This is possible 
because of the non-linearly realised supersymmetries, which have been
emphasized recently in \hamo ,\BL.  
 
Given the existence of the $(422)$ system, T-duality 
leads us to expect the existence 
of  $(420)$ bound states. 
This leads us to look for  supersymmetric solutions of $U(N)$
with magnetic flux in one 2-plane only.  
 We indeed find that known solutions of $SU(N)/Z_N$
combined appropriately with fields in the $U(1)$ 
do indeed yield $U(N)$ solutions with the 
right properties. 

T-duality also relates these systems 
carrying two-brane charges only.  These relate the  
constraints on the existence of torons \thtor, to geometrical  
 constraints on supersymmetric 
systems of intersecting branes at angles, which 
were first discussed in ref. \BDL .

While the instanton numbers in the $SU(N)/Z_N$ theory
can be fractional, only combinations of $SU(N)$ and $U(1)$ 
fields which have integral total  instanton number are solutions of 
the $U(N)$ gauge theory.
 This can be understood simply in terms of  Dirac quantization 
of D-brane charges.  

In section 5 we discuss the masses of these bound states, 
and their large N limit. We show, following arguments
in \witbd,  that the energy of the
lowest lying states in certain topological sectors
of  the  $SU(N)$ 
Yang Mills theory on $T^4\times R$ vanishes in this limit.

\newsec{ Preliminary remarks}  

\subsec{ Brane charges and fluxes.}
We will consider  $4$-branes aligned along the directions 
$(1234)$,  $2$-branes aligned along $(12)$, another 
 2-branes along $(34)$, and $0$-branes. 

 The low energy effective theory of 
  $N$ 4-branes  is $U(N)$ Yang Mills, 
and we will 
 look for the bound state by studying 
configurations in this theory.  
 The presence of $2$ branes along the $(12)$ 
direction  corresponds to  a $U(1)$  magnetic flux in the $(34)$ 
plane, and the presence of the $2$ branes along the 
 $(12)$ plane corresponds to the presence of magnetic flux  
along the $(34)$ plane.   
This follows from the Chern Simons couplings
of the field strengths to the RR potentials \docoup ,  
\eqn\cs{ \int C_{012} tr F_{34} + C_{034} tr F_{12}.  }  
A field strength in the $U(1)$ acts as a source for 
 the appropriate charge, but  this field strength 
automatically implies that we have `t Hooft flux in the 
$SU(N)$ part, see \witbd, \bsv , \senmarg. We will see this 
more explicitly below. 
Finally an  instanton embedded in the directions 
 $(1234)$ acts as  a source for the 0-brane. 
This follows from the 
interaction
\eqn\csii{ \int C_{0} tr F_{12} F_{34}.  }

The next step is to find  solutions of 
the $U(N)$ theory corresponding to 
supersymmetric configurations of branes. 
This requires embedding   't Hooft's solutions of 
$SU(N)/Z_N$ into  $U(N)$ gauge theory. 
As a preliminary to doing that  
we will discuss classical  solutions of the $U(1)$ theory.

\subsec{ $U(1)$ solutions  on a torus }  
 Consider the configuration 
\eqn\youone{\eqalign{ 
&A_1= B_{12} x_2 \qquad A_2=0 \cr
& A_3 = B_{34} x_4 \qquad A_4=0, \cr }}
on a torus with sides of lengths $a_1$, $a_2$,
$a_3$, and $a_4$.  
It satisfies non-trivial boundary conditions. 
Translation by $a_2$ in $x_2$ is accompanied by a gauge transformation 
 by $e^{ iB_{12} a_2x_1}$, and translation in $x_4$ is accompanied 
 by a gauge transformation $e^{ iB_{34} a_4x_3} $. 
Transporting a unit charge around the $12$ plane gives 
the quantization condition $B_{12} a_1a_2 = 2\pi n_{12}$, 
 whereas transporting in the $(34)$ plane gives the 
condition $B_{34} a_3a_4 = 2\pi n_{34}$.
Combining these two gives :
\eqn\comcon{ {a_1a_2 \over a_3a_4} = 
{n_{12}\over n_{34}} {B_{34}\over B_{12}} }
  If we 
 further impose the self duality $B_{12}= B_{34}$, 
we find that the ratio
of the box sizes is a rational number.  
However, self-duality of the $U(1)$ fields is not necessary 
for a supersymmetric configuration. As we will 
see in more detail later, requiring supersymmetry for 
$U(N)$ solutions will lead to similar constraints on box sizes.
This follows by taking into account both the linear
and nonlinear supersymmetries of the D-brane worldvolume 
theory, as given for example in \hamo: 
\eqn\supsym{ 
 \delta \lambda = \xi_1 \Gamma^{MN}F_{MN} + \xi_2 1  }
  The second term is proportional to the unit matrix  
in the Lie algebra, so \supsym\ allows  constant field strengths
in the $U(1)$, with no self-duality restriction, 
  to give BPS configurations. 
If the field strengths live in the $SU(N)$ part of the 
algebra, a generic constant (in spacetime) field will 
not be susy, but self-dual or anti-self dual fields  will
yield susy configurations.

In the non-abelian case there can be solutions with $A_{\mu}^a =0$ 
satisfying the twisted boundary conditions, as in first part 
 of \thtor. In the abelian case 
 this is not possible, see \ambfly.  This will also 
constrain the kind of supersymmetric solutions that 
can be constructed. 
 
\newsec{ Embedding solutions in $U(N)$ }

In discussing the $U(N)$ theory we will recall that 
there is a map from $SU(N) \times U(1)$ to $U(N)$. 
Let $(U,e^{i\theta})$ be 
an element of the product group. Then the map from
$SU(N) \times U(1)$ to $U(N)$ takes this to $e^{i\theta}U$ in 
$U(N)$. Notice that the elements $ (e^{2i\pi n \over N},
 e^{- 2i\pi n \over N})$ are in the kernel of this map. 
Therefore we can arrange the twists to be trivial in $U(N)$ 
by cancelling them  between $SU(N)$ and $U(1)$.  
 This requires consistently combining  solutions 
of $SU(N)/Z_N$ with   $U(1)$ solutions similar to 
those  described above, 
in such a way as to cancel the total twist.

One general remark about the instanton numbers of the solutions 
that we get in  this way can be made immediately.
Using the relations between the fields in the $U(1)$ and 
the twists,  we have:  
\eqn\uone{\eqalign{ 
&B_{\mu \nu}a_{\mu}a_{\nu} =  
{2\pi n_{\mu\nu} \over {N}} mod ~ (2\pi) \cr}}
This allows us to express the contribution to 
the  instanton number coming from the 
$U(1)$ fields  in terms of the twists:
\eqn\uon{ 
 {1\over {16 \pi^2}}  \int tr F_{\mu \nu} \tilde F_{\mu \nu} = 
{1\over N} n_{\mu \nu} \tilde n_{\mu \nu} + integer, } 
where $F_{\mu \nu} = B_{\mu \nu} 1$ is an $N\times N $ matrix. 
 The sign is important. The fractional part coming 
from the $SU(N)$ is also determined in terms of the twists
\thtor\ to be $-{1\over N} n_{\mu \nu} \tilde n_{\mu \nu}$. As a result
the total instanton number is integral. Since the instanton 
number is related to zero brane charge inside a 4-brane, 
we may interpret the integrality of the 
total instanton number as a consequence of the existence of the 
six-brane of ten dimensional 
type IIA string theory  with known charge, 
and Dirac quantization.

\subsec{   $(4220)$ solutions with trivial $SU(N)$ gauge fields.} 
These bound states of 4, 2, and 0-branes have the 
property that the element 
of $H^2(T^4, Z)$ defined by the $2$-brane charge 
has non-zero intersection number, but the total intersection number 
is zero.  There are solutions of the $SU(N)/Z_N$ theory 
 which have zero gauge fields in the presence of non-zero  
twists $n_{12}$,  $n_{34}$ where  
 ${n_{12}n_{34}\over N}$ is an integer.
We briefly review the construction of these solutions \thtor. 
Translation by $a_{\mu} $ in the $x_{\mu}$ direction 
is accompanied by a gauge transformation $\Omega_{\mu}$ 
which are constrained by  the equation : 
\eqn\omegcon{ \Omega_{\mu} (x_{\nu} = a_{\nu})
               \Omega_{\nu} (x_{\mu} = 0 ) = 
            \Omega_{\nu}(x_{\mu} = a_{\mu})
             \Omega_{\mu} (x_{\nu} = 0 ) e^{{ i\pi n_{\mu\nu} \over N}}. }  
For vanishing gauge fields there are 
solutions of the form 
\eqn\omegeq{ \Omega_{\mu} = P^{s_{\mu}} Q^{t_{\mu}}.  } 
where $PQ= QP e^{ {2i\pi\over N} }$.  
The equation \omegcon\ gives 
\eqn\esstt{ n_{\mu \nu} = s_{\mu} t_{\nu} - s_{\nu} t_{\mu} (\hbox{ mod N})} 
A complete characterization of these solutions requires 
specifying the appropriate set of $s$ and $t$ parameters. 
A necessary condition for this equation to be soluble is
\eqn\sol{ {1\over 8} \epsilon^{\mu \nu \alpha \beta} n_{\mu \nu} 
                    n_{\alpha \beta} = 0 {\hbox { mod N} }.    } 
This means that the above $U(1)$ 
 solutions can be simply embedded in the $U(N)$ theory, when the 
twists satisfy the above condition. 
 With the twist chosen as above, the instanton number 
is $n_{12}n_{34} \over N$. 
This gives a $(4220)$ system. 
Since  $SU(N)$ solution with non-trivial twists 
and  zero gauge fields 
only exist when the product $n_{12}n_{34}= 0 (mod~N)$
\thtor,  requiring $U(N)$ gauge invariance implies that the 
 zero brane charge can only be integral.  

This system has the charges N $4$-branes wrapped 
on $(1234)$, $n_{34}$ units of two-brane charge on the $(12)$ cycle,
$n_{34}$ units of  two-brane  charge on the 
on $(12)$ cycle, and  ${ n_{12}n_{34} \over N }$ units of 0-brane 
charge.  
 T-duality along the 1 and 3 directions
  gives a system with charges 
of two-branes in the  $(24)$, $(14)$, $(23)$ and $(13)$
planes. We will discuss this relation in more detail in section 4.

\subsec{ $(422)$  solutions with non-trivial 
$SU(N)$ gauge fields. }  

These solutions have the property that 
they have 4-branes and 2-branes with 
non-zero intersection number and yet 
no zero branes. If we use only 
$U(1)$ fields, there is necessarily 
a non-vanishing 0-brane charge related to the
intersection number of the 
$2$-branes. It is therefore important to turn 
on the $SU(N)$ fields to get configurations 
with these kinds of charges.

The following is  a construction of solutions with these 
charges.  We consider diagonal fields 
which break $U(N)$ to $U(k) \times U(l)$.
Each block has  vanishing $SU(l)$ or $SU(k)$ gauge potentials
and a solution may be obtained 
by taking two copies of the solutions of the
type obtained in the previous section.  
The $SU(N)$ field strengths however
are no longer vanshing.   There are now 
two sets of twists $n^{(l)}_{\mu\nu}$
and $n^{(k)}_{\mu\nu}$.  The total twist in 
the $U(1)$ sector, which corresponds
to 2-brane charge is given by 
\eqn\toobc{ 
n_{\mu\nu} = n_{\mu\nu}^{(l)} + n_{\mu\nu}^{(k)}. }
As before we will take all 
twists except those in the $(12)$ and $(34)$ planes
to vanish.  The 0-brane charge 
(instanton number) is then given by 
\eqn\instno{
	    P = P^{(l)} + P^{(k)}  
              = {n_{12}^{(l)}n_{34}^{(l)}\over l} + 
		{n_{12}^{(k)}n_{34}^{(k)}\over k}.
	    }
 
To get the $(422)$ system, we take  
an ansatz for the non-zero $U(N)$  fields of the form, 
\eqn\ans{\eqalign{  F_{12} = Diag ( B_{12} , 0) \cr
          F_{34} = Diag (  0 , B_{34} ). \cr }}
This corresponds to taking $n_{12}^{(k)}= n_{34}^{(l)} = 0$. 
Flux quantization conditions take the form: 
\eqn\fquan{\eqalign{ 
&B_{12} a_1a_2 = { {2\pi n_{12}^{(l)}}  \over l} \cr 
&B_{34} a_3 a_4 = { { 2\pi n_{34}^{(k)}}  \over k} \cr }}
All other twists $n_{\mu \nu}^{(k)}$ and $n_{\mu \nu}^{(l)}$ 
  are zero. 
Requiring the traceless parts 
$F_{ij} - {1\over N}tr F_{ij} $, to be self-dual
imposes $B_{12} = - B_{34}$.
These equations imply a condition on the   box sizes    
\eqn\boxcon{\eqalign{ 
   {a_{1}a_{2} \over {a_3 a_4}} =  
            { n_{12}^{(l)} k \over { n_{34}^{(k)} l} }. }} 
 The trace part does not 
need to be self-dual for supersymmetry to be preserved
as we see from \supsym. 
To make sure that these are acceptable 
solutions  we need to find integers 
$s_{\mu}^{(l)}, t_{\mu}^{(l)}$ which satisfy the condition
\eqn\stcon{ n_{\mu \nu}^{(l)} = s_{\mu}^{(l)}  t_{\nu}^{(l)} 
- s_{\nu}^{(l)} t_{\mu}^{(l)}  
             \hbox{ mod } l, }
for fixed $n_{\mu \nu}^{(l)}$ and $l$.   
This is easily solved  in this example, for instance take
$s_1^{(l)}= n_{12}^{(l)}$ and $t_{2}^{(l)}=1$ with other 
$s$, $t$'s set to zero.    

The instanton number coming from the 
$SU(N)$ part of the theory is 
$-{n_{12}^{(l)}n_{34}^{(k)} \over N}$ and the contribution 
 coming from 
the $U(1)$ part is the opposite. These two contributions
can in general be fractional.    
If we choose the twists $n_{(12)}^{(l)}= n_{(34)}^{(k)} $, 
the box sizes are directly related to the pattern of 
gauge symmetry breaking. 

This configuration then has the charges of  $4$-branes
aligned along directions $1,2,3,4$ and 2-branes
along $1,2$   and $2$-branes aligned along $34$. 
T-duality along $(12)$ direction can be performed giving 
a system with $(420)$ charges, which will be discussed further 
below. T-duality along $1$ and $3$ gives a system with 
charges of two-branes along $(24)$, $(14)$ and $(23)$.
This may be interpreted as a system obtained by two  
 two-branes whose projections 
 along $(24)$ are $k$ and $l$  respectively. 
One of them is  
rotated off the $(24)$ plane by rotations in the $(12)$. 
The other system is rotated off the $(24)$ plane 
by rotations in the $(34)$ plane.

\subsec{  $(4220)$  solutions with non-trivial 
$SU(N)$ gauge fields.}
 
 These bound states of 4, 2, and 0-branes have the 
property that the element 
of $H^2(T^4, Z)$ defined by the $2$-brane charge 
has non-zero intersection number, but the total intersection 
number does not have to be zero (unlike the case of section 3.1). 
Given the discussion in the last section, 
it is clear that there are $(4220)$ solutions 
with non-vanishing $SU(N)$ fields obtained by breaking 
$U(N)$ to $U(l) \times U(k)$. The fields live along the
diagonals of the $U(l)$ and the $U(k)$.

\subsec{ $(420)$ solutions }
These bound states have the property that the 
$2$-brane charge as an element of the 
$H^2(T^4, Z)$ has zero intersection number. 
The $(422)$ systems are T-dual to 
to $(420)$.  
The existence of the $(420)$ states suggests that we look 
for solutions which are BPS and which have instanton number in spite 
 of having a twist in only one plane. This would be  impossible 
 for a $U(1)$ theory but is  possible for $U(N)$ because of the 
non-trivial interplay between the $U(1)$ and the $SU(N)$ parts.
We need a twist in the $(12)$ plane  only, say. 
The key fact which makes this possible is that 
$SU(N)$ solutions can be constructed which have 
non-trivial twists in one plane only.

\newsec{Geometrical constraints}
The existence of the above solutions requires satisfying 
some striking constraints.  
For example for one class of solutions 
of $(422)$ type, the ratios of box sizes are 
rational, and the rational number in question 
is related to the structure of symmetry breaking 
caused by the solution. In this section we understand 
these constraints by relating them to the 
geometry and supersymmetry of systems of two-branes.

All the above systems could be mapped 
by appropriate T-dualities to systems 
carrying $2$-brane charge only. 
This suggests an interpretation in terms of 
$2$ branes possibly at angles.
The first type of solution we considered had 
no symmetry breaking and all the field 
strengths are constant in spacetime and live in the $U(1)$. 
They can be interpreted as T-duals of a single 
brane at an angle to the directions where T duality is 
performed.  
 The other solutions, which involve non-trivial 
 $SU(N)$ gauge fields  and which break 
the $U(N)$ to $U(k) \times U(l)$ 
 can  be  obtained by dualizing 
 a configuration of 2 sets of 
 two branes at a relative angle. since  we have two 
(sets of )  2-branes treated differently  the pattern of 
symmetry breaking is easily understood. The linear form of the gauge 
potentials as a function of the coordinates, is interpreted 
after T-duality in terms of rotations \polrev. The fact that 
the contribution to 
the  instanton number coming from each
block   is determined in terms of the 
fluxes corresponds to the condition which says that the two-brane 
configuration is not self-intersecting.  
The  integrality of the  net instanton number 
is also evidence that this picture is correct.   
 If we had fractional zero brane charge,  
in the T-dual system this would mean that there was 
fractional two brane charge in the system of two brane 
oriented at an angle, which would be a contradiction
{\foot{This point was developed in a discussion with S. Mathur. }}.

Having outlined the general arguments in favour 
of the relation between these solutions 
and systems of two-branes, we describe in
 some more detail how the constraints
match in two classes of examples. 

\subsec{ the $(4220)$ with $U(1)$ gauge fields only}

This system has the charges of $4$-branes wrapped 
on $(1234)$, two-branes wrapped on $(12)$, two-branes 
on $(34)$, and 0-branes. A T-duality along 
$(12)$ gives another $(4220)$ system. 
A T-duality along $(13)$ gives a system with charges 
of two-branes along $(24)$, $(14)$, $(23)$ and $(13)$
planes.  This can be interpreted in terms of a
system of $N$ two-branes which start from the 
configuration parallel to the $(24)$ plane and get rotated by 
angles $\theta_{12}$ and $\theta_{34}$ which mix respectively 
the directions $(12)$ and $(34)$. Since there is only one 
set of branes, we do not expect, from the worldsheet description,
any constraints from supersymmetry. The only constraints come from 
the requirement that the 2-brane charge is an integral combination 
of the 2-cycles  of the torus. These correspond to the 
flux quantization conditions in the Yang Mills side. 
 The absence of any extra constraints coming from the 
susy corresponds to the fact that the field strength 
does not have to be self dual to guarantee BPS saturation.

Given a two brane at  an angle, it is clear that 
we can increase all the charges by a factor 
S by just having a stack of S two-branes parallel 
to each other. In Yang Mills this requires that 
given any solution with  fluxes, instanton 
number and $s_{\mu}$, $t_{\mu}$ etc. specified, we should be 
able to scale up $N$, the fluxes and the instanton number 
by the factor $S$, and recover a solution with the new 
parameters. This is indeed possible, 
if we accompany it by a scaling of 
$s_{\mu}$ by a factor of $S$, leaving the $t$ fixed.

A map of the two brane to the target space is given
by 
\eqn\map{ 
	 X^{\mu}(p,q) = b^{\mu}m^{\mu}_p p + b^{\mu}m^{\mu}_q q,
	}
where $p$ and $q$ are defined on the interval $[0,1]$,  and
$b_{\mu}$ are the dimensions of the  torus related by T duality 
along  $1$ and $ 3$ to the torus containing the $(4220)$ system.
The 2-brane charges $Q_{\mu\nu}$ are equal to the covering numbers in
the $\mu\nu$ plane:
\eqn\cov{
	Q_{\mu\nu} =  (m^{\mu}_p m^{\nu}_q - m^{\mu}_q m^{\nu}_p).} 
 Because these charges are related to 
covering numbers through~\cov,  they satisfy the constraint 
\eqn\const{
	    \epsilon_{\mu\nu\alpha\beta}Q_{\mu\nu}Q_{\alpha\beta} = 0.
	  }	
This is just the condition that the two-cycle defined 
by the surface of the two-brane is not self intersecting. 
After undoing the T duality in the $1$ and $3$ directions, 
this corresponds to constraints in Yang Mills theory
relating the instanton number to the fluxes and the rank of the gauge 
group.   
For instance in the case in which the $SU(N)$ field strength vanishes, 
and there are non-vanishing $U(1)$ fluxes $n_{12}$ and  $n_{34}$, 
the instanton number is given by $P = n_{12} n_{34}/N$.
After T duality in the $1$ and $3$ directions this corresponds to 
$Q_{13} = Q_{14}Q_{23}/Q_{24}$,  which is equivalent to the 
constraint \const.

\subsec{ the $(422)$ system}

The geometrical system which is T-dual
 (by a T duality along the  directions
$1$ and $3$)  to 
the class of solutions with $(422)$ charges which we 
described in section 3,  consists of 
a brane  $B^{(l)}$ (corresponding to $U(l)$ )   which 
lies along the 2-cycle $Q_{24}^{(l)} (24) + Q_{14}^{(l)} (14) $
and a brane $B^{(k)}$ (corresponding to  $U(k)$) 
 which lies along the 2-cycle 
$Q_{24}^{(k)} (24) + Q_{23}^{(k)} (14) $.
The geometric constraint \const\ is separately satisfied in each 
unbroken subgroup. This can be seen from \instno .
  The 2-brane $B^{(l)}$ corresponding to 
$U(l)$ subgroup has the nonvanishing charges
\eqn\charges{\eqalign{
		Q^{(l)}_{14} = n^{(l)}_{12},\cr
		Q^{(l)}_{24} = l.\cr }}
The 2-brane $B^{(k)}$ corresponding to the $U(k)$ 
subgroup has non-vanishing charges : 
\eqn\chargesk{\eqalign{
		Q^{(k)}_{23} = n^{(k)}_{34},\cr
		Q^{(k)}_{24} = k.\cr }}
The  brane  $B^{(l)}$  has been rotated off
the $(24)$ plane by an angle $\theta_{12}$ in the 
$(12)$ plane 
\eqn\tangl{ 
 tan \theta_{12} = { Q_{14}^{(l)} b_1 \over { Q_{24}^{(l)} b_2 } }}
The  brane  $B^{(k)}$  has been rotated off
the $(24)$ plane by an angle $\theta_{34}$ in the 
$(34)$ plane 
\eqn\tangk{ 
 tan \theta_{34} = { Q_{23}^{(l)} b_3 \over { Q_{24}^{(k)} b_4 } }}
Using the T-duality relations 
(in units where $4\pi (\alpha^{\prime})^2 = 1$):  
\eqn\tdu{\eqalign{ 
& b_1 = {1\over a_1},  b_2 = a_2 , b_3 = {1\over a_3}, b_4 = a_4  \cr 
& 2\pi ~  tan \theta_{12} = B_{12}, 2\pi ~  tan \theta_{34} = B_{34},
 \cr } } 
we see  that the geometrical equations \tangl\ and 
\tangk\ are  precisely the flux quantization conditions in
 \fquan . 

It follows by considerations starting from the 
worldsheet formulation of 2-branes \BDL\ that 
such a system with $\theta_{12}= \theta_{34}$ 
is  supersymmetric. Therefore  the pair 
of branes can be supersymmetric and  
compatible with the 
 periodicities defining the torus,  if the box sizes satisfy the 
constraint which is precisely the one in \boxcon .

\newsec{ Mass formula and quantum 
ground states  of $SU(N)$ Yang Mills} 

 The supersymmetry algebra 
can be used to obtain the masses of the 
bound states with charges $(4220)$.
 It has been obtained in \gilad\ by considering the 
exchange of gravitons between the bound states. 
 We can also obtain  
 it by relating it to a T-dual system 
of minimal area 2-branes.
The rotated $2$ brane has an area which is given by : 
 \eqn\area{ A^2 = ( (Q_{24} b_2b_4)^2 + ( Q_{14} b_1b_4)^2 + 
                   ( Q_{23} b_2b_3)^2 + (Q_{13} b_1b_3)^2  )  } 
In our  units ($4\pi (\alpha^{\prime})^2 = 1$)
 all the brane tensions are equal, and gauge fields 
under T-duality are  $2\pi$ times the corresponding coordinates. 
The mass of the two-brane  can then be expressed in terms 
of the above data and the string coupling. 
Under this T duality, $Q_{24}$ becomes the number of 4-branes, i.e 
$N$ in the gauge theory, $Q_{14}$ becomes two brane number in the 
$(34)$ plane or $n_{12}$ in the gauge theory language, 
 $Q_{23}$ becomes 2-brane charge in the $(12)$ plane or 
the flux number $n_{34}$ in the gauge theory language; and finally 
$Q_{13}$ becomes the zero brane charge or instanton number. 
The box sizes are related as in \tdu. 
Recalling the 
transformation of the string coupling under T duality 
 we  can express the mass in terms of the 
charges of the $(4220)$ system. 
\eqn\mas{ m^2 = ( Q_4^2 + (Q^{(12)}_2)^2 + (Q^{(34)}_2)^2 + Q_{0}^2 ).  }

If we consider fluxes such that $n_{12}n_{34} /N$ is of order 
$1$, and we expand in large N, 
 then the above expression simplifies to an expression 
with three terms which contain the two-brane and 4-brane charges 
quadratically. The term coming from the instanton number 
only affects higher orders in the $1/N$ expansion. 
The contribution from the classical Yang Mills action
for the $U(1)$ part  
correctly gives the excess energy of the two-branes. 
This means that the contribution 
 from the $SU(N)$ sector is zero. This gives a non-trivial 
prediction for the energies of states of minimal energy 
in  supersymmetric ( with 16 supercharges) 
 $SU(N)$ Yang Mills theory on $T^4 \times R$  with 
fixed box sizes, and fluxes ${ n_{12} n_{34} \over N}$
of order 1, as $N$ goes to infinity.    
This is the same kind of argument that was used in 
\witbd\ and \senmarg\ to deduce properties of supersymmetric 
Yang Mills theories. In cases considered 
in \senmarg,  where only an electric 
flux in some definite direction in the torus was present, 
the questions about ground states of Yang Mills 
on tori could be reduced to lower dimensional questions 
because fluctuations in directions transverse to the direction 
of the flux could be ignored. The questions we are considering here  
 cannot be reduced to questions of quantum mechanics 
or of $1+1$ dimensional field theory, 
so they probe more  ``higher dimensional'' properties 
of the 5-dimensional Yang Mills theory (regulated by its  
embedding in the theory of 4-branes).  A better understanding 
of these properties is  necessary in the context of 
the matrix model approach of  \BFSS\ and the approach to 
compactification explored in \BFSS\wati\sus\grt.

\newsec{ Comments and conclusions.}

In the above we have discussed in detail 
supersymmetric solutions where $U(N)$ is 
left unbroken, and the case where it is broken to 
$U(k)\times U(l)$. The former were related to systems 
with one  (stack of) non self-intersecting branes
 aligned along some 
2-cycle in $T^4$. The latter were related to two ( stacks  of ) 
branes at a relative angle. The solutions may be generalized 
to situations where the unbroken symmetry group 
has more than two factors. Correspondingly on the two-brane side, 
there are solutions with more than two sets of branes  
which preserve supersymmetry \BDL.

T duality on all $4$ directions of the $T^4$ gives relations 
between instanton moduli spaces of $U(N)$ of instanton 
number $k$ and instanton moduli spaces of $U(k)$ of instanton 
number $N$ (together with a reshuffling of two brane fluxes), 
as mentioned in the physics literature in \bsv\ and \domo. 
 In most of the discussion relating two-branes to 
the systems of 4-branes we T-dualized in a 
definite pair of directions, namely the directions 1 and 3. 
As a consequence the two-brane charges in the $(24)$ plane
determined the ranks of the gauge groups involved. 
If we started from the same two-brane system and T-dualized 
in the directions 2 and 4, the ranks of the gauge groups 
would be related to two brane charges in the $(13)$ plane. 
This suggests that the symmetries which exchange 
instanton number and rank of gauge group 
can be made manifest by considering 
systems of 2-branes that they are dual to. 

One of our motivations for putting the 
torons of \thtor\ in the context of D-brane bound states
was to understand the fractional tension strings appearing
in the effective string models of black holes 
\malsus\mal\malstrom\cgkt\km . The systems we have considered
in this paper are simpler and have more supersymmetry than those
of interest in these papers, but some qualitative similarities 
between the behaviour of fractional instantons in the systems 
we discussed and those of fractional strings entering black hole
models may be noticed. In both cases the fractional objects 
are useful building blocks but 
 enter the physical systems in combinations 
which have integral net charge. In the black hole
context only the integrally charged objects can interact 
with closed strings. In the systems we studied 
imposing the full $U(N)$ gauge symmetry required that the 
net instanton charge be integral. Similar behaviour can be seen 
in another simple system with fractional branes
studied in \dm\ and \akfrac.  
 
Another comment may be made concerning   
possible relations to black holes. As emphasized in 
\malstromii,  the effective string models only use  
 the tree level string theory or the conformal field theory. 
We have seen in section 5 that the systems we studied have simple 
behaviour at large N, the classical $U(1)$ energy 
agreeing with the correct BPS formula in appropriate flux sectors. 
This suggests that  the surprising effectiveness of the tree level 
effective string model may be related to  
 the ``classical'' behaviour of large $N$ gauge 
theories.

\bigbreak\bigskip\bigskip
\centerline{\bf Acknowledgments}\nobreak
We wish to thank O. Ganor, J. Maldacena,  S. Mathur  and W. Taylor 
for helpful discussions. This research  is supported by NSF Grant
PHY96-00258.

\listrefs
 
\end